\documentstyle[12pt,epsfig,epsf]{article}

\newcommand{\be}{\begin{eqnarray}}
\newcommand{\en}{\end{eqnarray}}
\newcommand{\ph}{\phantom{00}}

\def\biggg#1{{\hbox{$\left#1\vbox to 20.5pt{}\right.\n@space$}}}

\def\Biggg#1{{\hbox{$\left#1\vbox to 23.5pt{}\right.\n@space$}}}

\title{                                                      
Numerical simulations of a two-dimensional \\
lattice grain boundary model \\[8mm]} 
\author{                                                                        
A.~Jaster, H.H.~Hahn \\[4mm] 
Institut f\"{u}r Theoretische Physik, TU Braunschweig, \\                          
Mendelssohnstr.\,3, D-38106 Braunschweig, Germany}           
                                                                                
\date{May, 1997}

\begin{document}

\maketitle
\begin{abstract}

We present detailed Monte Carlo results for a  two-dimensional grain
boundary model on a lattice. The effective Hamiltonian of the system
results from the microscopic interaction of grains with
orientations 
described by spins of unit length, and leads to a
nearest-neighbour interaction proportional to the 
absolute value of the angle between the grains.
Our analysis of the
correlation length $\xi$ and susceptibility $\chi$ 
in the high-temperature phase
favour a Kosterlitz-Thouless-like (KT)
singularity over a second-order phase transition. Unconstrained
KT fits of $\chi$ and $\xi$ 
confirm the predicted value for the critical exponent $\nu$,
while the values of $\eta$
deviate from the theoretical prediction.
Additionally we apply finite-size scaling theory and
investigate the question of multiplicative logarithmic corrections
to a KT transition. As for the critical exponents 
our results are similar to data obtained from
the XY model, so that  
both models probably lie in the same universality class.

\end{abstract}


\section{Introduction}

Phase transitions in two-dimensional systems with continuous 
symmetry are of great interest since many years. 
In these systems conventional 
long-range order at non-zero temperature cannot occur \cite{MERWAG}.
However, two-dimensional particle systems or the XY model 
are characterized 
at low temperatures by  quasi-long-range order, while the high-temperature
phase is disordered.
The critical behaviour of the models 
depends on the dimension of the order parameter. For
the two systems mentioned the order parameter dimension is two. 
Physical realizations of two-dimensional $O(2)$ symmetric 
systems are films of liquid
helium or superconductive layers.

The XY model provides a simple model of a system with 
continuous symmetry. For 
the phase transition numerical
studies favour a Kosterlitz-Thouless-like  \cite{KOTTHO}  behaviour.
The KT mechanism is based on a topological defect (called vortex)
unbinding scenario. In the low-temperature region vortices 
are bound in pairs.
The dominant excitations are spin waves, which destroy longe-range order.
Increasing the temperature leads to an unbinding of vortex pairs.
The theory predicts a continuous transition from the quasi-long-range 
phase to the disordered phase derived from renormalization group treatment.

Particle systems are characterized by two order parameters 
related to different topological defects.
Therefore, for the melting transition  additional complications arise.
There are several theoretical approaches for this transition.
Halperin and Nelson \cite{HALNEL} as well as Young \cite{YOUNG}
enhanced the ideas of Kosterlitz and Thouless. 
The KTHNY theory deals with a dislocation unbinding mechanism,
which is responsible for the melting transition, and 
a disclination transition,
which destroys the nearest-neighbour-bond orientation.
An alternative scenario has been proposed by Chui \cite{CHUI}.
He presented a theory via spontaneous generation of grain boundaries,
i.e.\ collective excitations of dislocations. He found that
grain boundaries may be generated before the dislocations unbind
if the core energy of dislocations is sufficiently small, and
predicted a first-order transition. 

Another possibility to study two-dimensional melting is
the simulation of the defect system itself. Saito performed Monte
Carlo simulations of dislocation vector systems \cite{SAITO}
and found that for systems with large dislocation core energy 
unbinding of dislocation  pairs causes a continuous phase transition, while
small core energies produce a first-order transition by 
formation of grain boundaries.
 
Patashinskii \cite{PATASH}
proposed a model, which can be seen as a mesoscopic lattice
model for grain boundaries. 
The energy per unit length
of a boundary between 
two grains  with small angle  
$\Delta\phi = \phi -\phi'$ is proportional to the 
absolute value of $\Delta\phi$.  
A simple ansatz for the description of the interaction 
of grains is then given by putting them 
on a square lattice with nearest-neighbour interaction
proportional the absolute value 
 for small lattice gradients. The orientation of
 the grains is described by spins of unit length
$\vec{s}({\rm\bf x})=(\cos \phi({\rm\bf x}),\sin \phi({\rm\bf x}))$.
Therefore Patashinskii used a Hamiltonian 
of the form
\be
H=\beta \sum_{<{\rm\bf x\, x'}>} \, \left | \, \sin 
\frac{N}{2} \left
 ( \phi({\rm\bf x}) -\phi({\rm\bf x'}) \right ) 
\, \right | \ .
\en
The number $N$ is related to the symmetry of the system.
For example, particle systems with a hexagonal symmetry
are described by $N=6$.
However, the partition function is essentially independent of $N$. 
Therefore we neglected this parameter and used angles 
of  range $0 \leq \phi({\rm\bf x}) < 2\pi$.
Also, we used the simplified model 
\be
H=\beta \sum_{<{\rm\bf x\, x'}>} \, \left |\,  \phi({\rm\bf x}) -
\phi({\rm\bf x'})
\,\right | \ , \ph -\pi< 
\phi({\rm\bf x}) -\phi({\rm\bf x'}) \leq \pi 
\en
for numerical simulations,
since (at least in the case of a continuous phase
transition)  only the behaviour for small $\Delta\phi$'s
is relevant. 
This coincides with  an XY model with the negative cosine  
replaced by its absolute value. 
In this paper we 
try to determine the order and mechanism of
the phase transition of this model.

The KT scenario predicts an exponential singularity for
the correlation length 
\be
\label{KTxi}
\xi(t)&=&a_{\xi} \, \exp \left ( b_{\xi}\, t^{-\nu}  \right ) \ , 
\en
and the susceptibility  $\chi$
\be
\chi(t)&=&a_{\chi} \, \exp \left ( b_{\chi}\, t^{-\nu} \right )
\en
if $t=\beta_{\rm c} -\beta \searrow 0$, while the specific heat $C_{\rm V}$
should not show any divergent behaviour. The critical exponent $\eta$
defined by
\be
\label{chixi}
\chi&\sim&\xi^{2-\eta} \ ,
\en
and the critical exponent $\nu$ are given by
\be
\nu=\frac{1}{2} \ , \ph \eta=\frac{1}{4} \ ,
\en
while $b_\xi$  is a non-universal constant and 
\mbox{$b_\chi=(2-\eta)b_\xi$}.
An alternative approach is a conventional second-order behaviour
with power-law singularities of 
\be
\label{PLxi}
\xi(t)&=&a_{\xi} \, t^{-\nu}
\en
and
\be
\chi(t)&=&a_{\chi} \, t^{-\gamma} \ .
\en
In the following we analyze our extensive Monte Carlo results
to answer the question of kind and order of the phase transition
of the grain boundary model.


\section{Simulations}

\subsection{Algorithms and measurement}

For the simulations we used a combination of the Metropolis,
over-relaxation \cite{OVERRELAX1,OVERRELAX2} and 
cluster \cite{CLUSTER1,CLUSTER2} algorithm, 
depending on the temperature and 
size of the lattice. The study was carried 
out on lattice sizes of $L^2=60^2$, $120^2$, $240^2$, $480^2$ 
and $960^2$ with periodic
boundary conditions.

For each simulation we measured the energy, specific heat, 'magnetization',
susceptibility, fourth-order cumulant and the 
zero momentum correlation function
(for a definition see eq.\ (\ref{eqSPINCORFCT}) below). 
In doing so, we used  conventional estimators for all observables.
Additional improved estimators \cite{IMPEST}
were measured for the  susceptibility
and correlation function. Typically, errors were of the 
order of $1\%$ or less.

For the calculation of the specific heat 
\begin{figure}[t]
\vspace{-0mm}
\begin{center}
\centerline{\epsfysize=11.0cm
\epsfxsize=12.0cm
\epsfbox{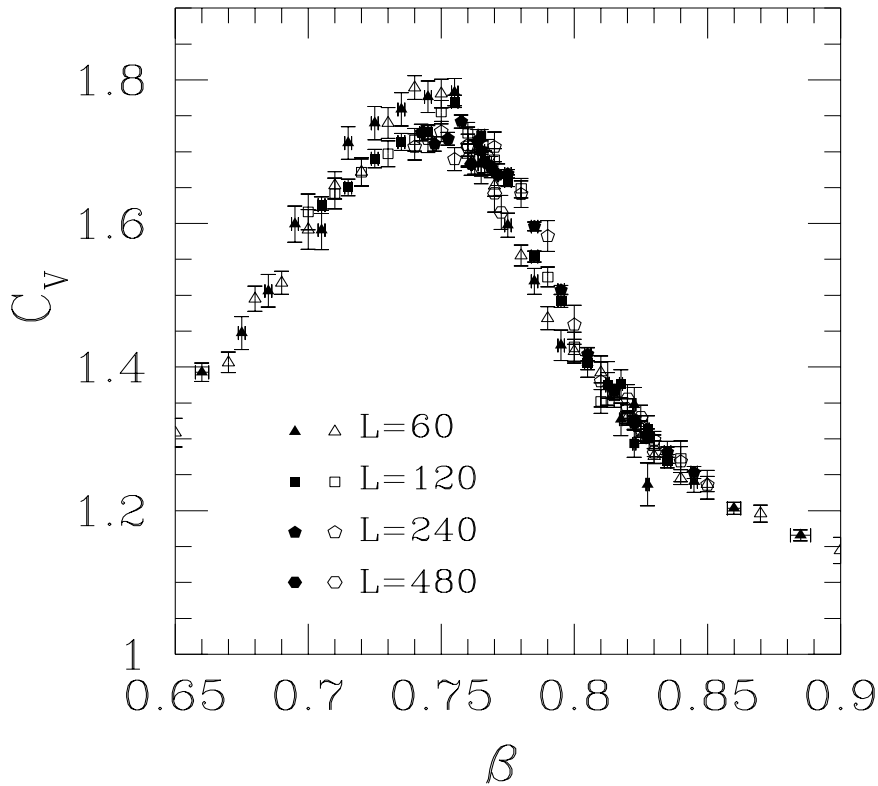}}
\parbox{13.5cm}{\caption{\label{FIG_SpecHeat}
Specific heat as a function of $\beta$.
Full symbols stand for the calculation by 
numerical differentiation, open symbols denote
the measurement of energy fluctuations.
}}
\end{center}
\end{figure}
we used two different ways: on  one hand we measured the fluctuations 
in the energy per point 
\be
C_{\rm V}=\beta^2\, L^2( \langle e^2 \rangle - \langle e \rangle ^2)
\en
 on the other hand we compute 
the derivative 
\be
C_{\rm V}=-\beta^2\,\partial e/\partial \beta \ . 
\en
We find that both
methods give consistent results as shown in fig.~\ref{FIG_SpecHeat}. Except
for the smallest lattice the data do not show any significant 
finite-size effects. As one can see the specific heat develops a smooth
peak at an inverse temperature below the transition point
$\beta_{\rm c}\approx 0.83$. At the transition point
the energy and the specific heat stay finite and show no divergent behaviour.
This is consistent with a KT-like transition and makes a first- or
second-order  transition unlikely.

The magnetic susceptibility was calculated from fluctuations 
in the magnetization
per point
\be
\chi=L^2 \, \langle  \vec{m}^2 \rangle
\ , \ph \vec{m}=\frac{1}{L^2} \sum_{x,y} \vec{s}(x,y)
\en
in the conventional manner and using improved estimators.
Finite-size effects were studied by comparing data on different lattice
sizes. We find that for correlation lengths  $\xi<L/7$ no   
lattice size dependence
of $\chi$ within the statistical errors can be seen.
 
In the high-temperature phase
the correlation length was extracted from the 'zero momentum' spin-spin
correlation function
\be
 \label{eqSPINCORFCT}
 \Gamma_{\vec{s}}(x) = \, \langle \vec{s}(x) \cdot \vec{s}(0) \rangle \ , \ph 
 \vec{s}(x)=\frac{1}{L} \sum_y \vec{s}(x,y)
\en
by fitting the data with a single $cosh$ or $cosh+const$
in the interval $x_0\leq x \leq L/2$. Most of the time,
constant contributions were  negligible,
because they are of order $exp(-L/\xi)$. To determine the 
influence of 'excitations' 
(smaller eigenvalues of the transfer matrix), we
compared the results for different minimal distance $x_0$. The
correlations are always dominated by the lowest state of
the transfer operator 'Hamiltonian',
so that we had to omit only few points.
As for the susceptibility we used  conventional and  improved estimators. 
The latter have substantially smaller statistical errors, because only 
positive terms contribute
to the correlation function. For $\beta>\beta_{\rm c}$ it is 
expected that the long 
distance behaviour of the spin-spin correlation function 
is given by a power-law
decrease with a critical  exponent $\eta$ which is a function 
of $\beta$, but we made no 
investigations to this point.

\subsection{Numerical results in the high-temperature phase}

In the following we will analyze the critical 
behaviour of $\xi$ and $\chi$.
Therefore, we have performed least square fits with
three different forms (according
to eqs.~(\ref{KTxi}) and (\ref{PLxi})). In the first case we 
used a four-parameter Kosterlitz-Thouless fit
(KT4), i.e.
\be
\ln(\xi)=a+b\,t^{-\nu} \ , 
\en
with $\beta_{\rm c}=\beta + t$ as the fourth parameter.
In the second case the value of $\nu$
was fixed to $0.5$ (KT3). Assuming a power-law behaviour of the divergence 
we obtained the third case: 
\be
\ln (\xi)=a - \nu \, \ln (t) \ .
\en
Systematic errors had been estimated
by varying the range of points to be fitted and by replacing 
$t=\beta_{\rm c}-\beta$ by $t=T-T_{\rm c}$. We used data 
subsets consisting of all points with $6<\xi < L/7$ and 
$12<\xi<L/7$, respectively\footnote{There is one exception:
for the point $\beta=0.7725$ we also got $L/\xi\approx
480/69.2 \approx 6.94$.}. This corresponds to
$0.67 \leq \beta \leq 0.78$ (20 points)
and $0.72 \leq \beta \leq 0.78$ (15 points).
Errors for the fitted parameters were computed by performing fits 
on data sets to a Gaussian
distribution of $\xi(\beta)$ with variance $\Delta \xi(\beta)$.
All calculations were performed  correspondingly for $\chi$.
\begin{figure}[t]
\vspace{-0mm}
\begin{center}
\centerline{\epsfysize=11.0cm
\epsfxsize=12.0cm
\epsfbox{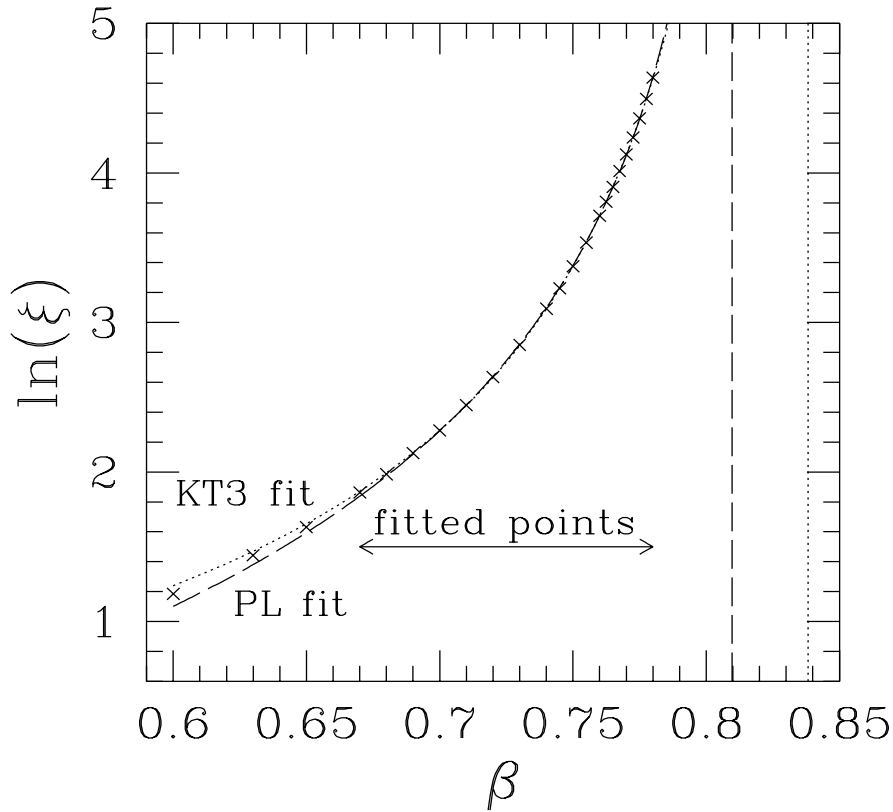}}
\parbox{13.5cm}{\caption{\label{FIG_KTPL1}
The curves shown are the best fit for a Kosterlitz-Thouless 
behaviour with fixed $\nu$ (dotted)
and a power-law ansatz (dashed). We
used all correlation lengths $\xi>6$. Statistical 
errors are too small
for a visualization. The critical values
of $\beta$ are visualized by  vertical lines. 
}}
\end{center}
\end{figure}

\begin{table}[t]                                                                  
\begin{center}                                                                  
\parbox{13.5cm}{\caption{ \label{KTvsPL}                                
Best fit parameters for the critical behaviour of the correlation length
and susceptibility. For $\xi>6$ we have fitted 20 points in 
\mbox{$0.67 \leq \beta \leq 0.78$}. In the case $\xi>12$ we used 15 points
in the range \mbox{$0.72 \leq \beta \leq 0.78$}.
}}                                                                              
\end{center}  

{\small                                                                  
\begin{center}                                                                  
\begin{tabular}{|cc||*{3}{r@{.}l r@{.}l|}}                             
\hline                                                 
\hline
 \multicolumn{2}{|c||}{\ } & \multicolumn{4}{c|}{KT4} 
&   \multicolumn{4}{c|}{KT3} &   \multicolumn{4}{c|}{PL} \\          
  \multicolumn{2}{|c||}{\ } 
 & \multicolumn{2}{c}{$\xi>6$} & \multicolumn{2}{c}{$\xi>12$} & 
   \multicolumn{2}{|c}{$\xi>6$} & \multicolumn{2}{c}{$\xi>12$} & 
   \multicolumn{2}{|c}{$\xi>6$} & \multicolumn{2}{c|}{$\xi>12$} \\
\hline
& ln(a)     &                                                                      
-3&{62(33)}   &   -2&{92(88)}   &  -2&{08(1)}   &  
-2&{01(2)}   &   0&{176(1)}   &  0&{148(2)}  \\                 
 & b     &                                
2&{88(28)}   &   2&{30(73)}   &  1&{62(1)}   &  
1&{59(1)}   &   \multicolumn{4}{c|}{}  \\
 $\!\xi(\beta)\!\!$ & $\beta_{\rm c}$     &                    
0&{8295(12)}   &   0&{8323(36)}   &  0&{8382(2)}   &  
0&{8373(3)}    &   0&{8097(1)}   &  0&{8118(2)}  \\    
 & $\nu$     &                                  
0&{350(33)}    &   0&{403(71)}    &  0&{5}   &  
0&{5}         &   1&{816(3)}   &  1&{901(6)}  \\                 
 & $\chi^2$/{\small dof}    &                                       
1&{10}    &   1&{17}    &  3&{37}   &  
1&{22}   &   28&{41}   &  6&{59}  \\                 
\hline                                                                    
& ln(a)     &                                                                      
-1&{45(13)}   &  -1&{25(40)}   &  -1&{32(1)}   &  
-1&{32(2)}  &   0&{713(2)}   &  0&{594(4)}  \\      
 & b     &                                                                      
1&{84(11)}   &   1&{67(33)}   &  1&{73(1)}   &  
1&{73(1)}   &   \multicolumn{4}{c|}{}  \\                
$\!\xi(T)\!\!$& $\beta_{\rm c}$     &               
0&{8335(13)}   &   0&{8353(34)}   &  0&{8348(1)}   &  
0&{8347(2)}   &   0&{8057(1)}   &  0&{8086(2)}  \\      
& $\nu$     &                                       
0&{480(19)}   &    0&{510(62)}   &  0&{5}   &  
0&{5}         &    1&{557(3)}   &  1&{669(5)}  \\                 
& $\chi^2$/{\small dof} &                 
1&{00}   &   1&{16}     &  1&{02}   &  
1&{07}   &   80&{64}   &  14&{08}  \\    
\hline                                                                    
& ln(a)     &                                                                      
-3&{7(12)}    &   -1&{0(20)}   & -2&{46(4)}   &  
-2&{32(10)}   &   0&{187(4)}   &  0&{126(9)}  \\                 
& b     &                                                                      
3&{7(10)}   &   1&{7(15)}   &  2&{75(2)}   &  
2&{69(5)}   &  \multicolumn{4}{c|}{} \\                 
$\!\chi(\beta)\!\! $& $\beta_{\rm c}$     &                                        
0&{8318(43)}   &   0&{8335(93)}   &  0&{8372(4)}   &  
0&{8361(9)}   &   0&{8069(3)}   &  0&{8099(6)}  \\                 
& $\nu (\gamma)$     &                 
0&{42(7)}   &   0&{64(17)}   &  0&{5}   &  
0&{5}         &   2&{974(12)}   &  3&{169(34)}  \\       
& $\chi^2$/{\small dof} &    
0&{94}   &   1&{25}   &  0&{97}   &  
1&{27}   &   4&{84}   &  2&{49}  \\                 
\hline        
& ln(a)     &                                       
-0&{8(5)}   &    0&{5(15)}   &  -1&{12(3)}   &  
-1&{13(8)}   &   1&{98(2)}   &  1&{33(5)}  \\                 
& b     &                                                                      
2&{6(4)}   &   1&{6(12)}   &  2&{91(2)}   &  
2&{91(4)}   &    \multicolumn{4}{c|}{} \\                 
$\!\chi(T)\!\! $& $\beta_{\rm c}$     &                            
0&{8359(40)}     &   0&{8455(99)}   &  0&{8332(4)}   &  
0&{8333(8)}   &   0&{8024(3)}   &  0&{8066(6)}  \\                 
& $\nu (\gamma)$     &                                        
0&{54(6)}     &   0&{71(27)}   &  0&{5}   &  
0&{5}         &   2&{524(9)}   &  2&{771(26)}  \\     
& $\chi^2$/{\small dof}   &                                        
0&{93}   &   1&{25}     &  0&{91}   &  
1&{23}   &   11&{88}   &  3&{06}  \\                 
\hline                   
\hline
\end{tabular}  

\end{center} }                                                                   
\end{table}     
Our results are summarized in table \ref{KTvsPL}. Additionally one 
example is plotted in fig.~\ref{FIG_KTPL1}.
One observes that the KT like fits look consistent 
(except one of the KT3 fits for $\xi(\beta)$)
and that both types  always give a similar $\chi^2$ per
degree of freedom (dof). The influence of the lower
bound of the fitted interval is  small.
This situation changes for the power-law fits.
The different intervals yield large changes in the fitted parameter.
Also, PL fits have always larger $\chi^2$/dof than the KT fits.
The data for the correlation length make a power-law 
divergence very unlikely. 

Taking $\nu$ as a free parameter (KT4) results in large
errors. This situation is similar as in the XY model 
\cite{GUPBAI,JANNAT}. The reason is that in the four-parameter space
a valley for $\chi^2$ exists, which is narrow in three dimensions, but 
flat in the fourth. The different fits
with $\nu$ as a free parameter yield the estimate 
\be
\nu=0.46(7) \ .
\en
Holding $\nu$
or $T_{\rm c}$ fixed leads to a clear minimum and 
much smaller errors in the remaining parameters.
From the $\beta_{\rm c}$'s of the three-parameter 
Kosterlitz-Thouless fits and the power-law fits
we got the following values:
\be
\label{betacKT}
\beta_{\rm c,KT3} = 0.8354(20) \ ,
\en
\be
\label{betacPL}
\beta_{\rm c,PL} =0.8079(30) \ .
\en

We further analyzed our data by investigating the relation between $\chi$ 
and $\xi$. To check the validity of eqn.~(\ref{chixi}) we
plot $ln(\chi/\xi^{7/4})$ versus $ln(\xi)$. 
\begin{figure}[t]
\vspace{-0mm}
\begin{center}
\centerline{\epsfysize=10.0cm
\epsfxsize=11.0cm
\epsfbox{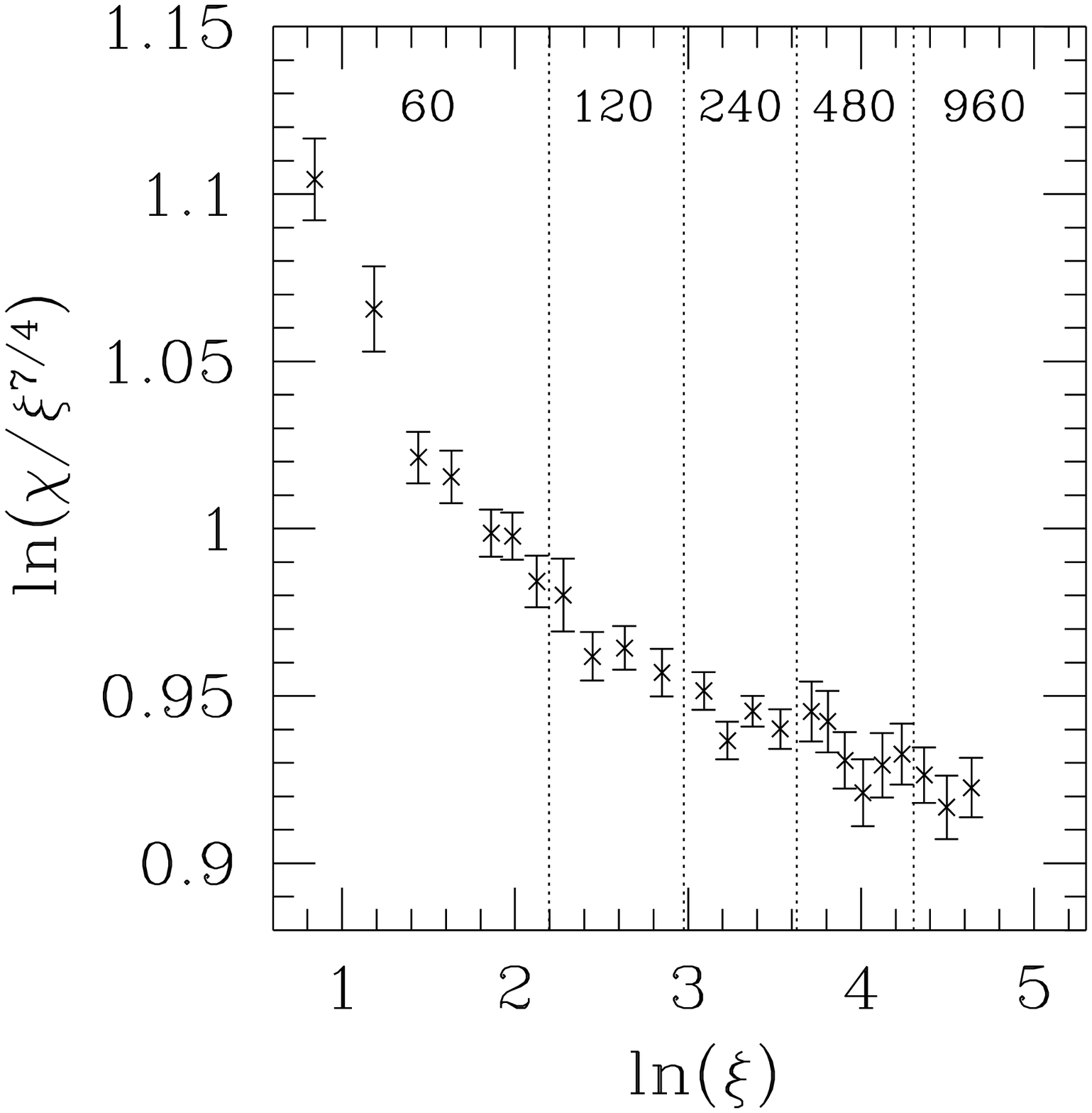}}
\parbox{13.5cm}{\caption{\label{FIG_eta1}
Test of the scaling relation $\chi \sim \xi^{7/4}$. The 
tendency to decrease implies $\eta>1/4$.
}}
\end{center}
\end{figure}
For the predicted KT value $\eta=1/4$ we should see a horizontal line.
A different value of $\eta$ would correspond to a straight line of 
non-zero slope. Indeed, fig.~\ref{FIG_eta1} shows  a 
negative slope, but with decreasing
absolute value for increasing $\xi$. Performing straight line fits for the 
points $\xi > 12$ results in an estimate of 
\be
\eta=0.270(3) \ . 
\en
For the points $\xi>50$ we got $\eta=0.259(17)$.
The  difference to the theoretical value can 
perhaps be explained by logarithmical 
corrections as discussed later. 

It should be noted that 
the values of the critical exponent
$\nu=0.48(10)$ and $\eta=0.267$ 
of the XY model in Villain's formulation 
\cite{JANNAT} (calculated with same methods)
coincide with our grain boundary model within statistical errors.
Also,  the behaviour
of $ln(\chi/\xi^{7/4})$ as a function of $ln(\xi)$ 
(fig.~\ref{FIG_eta1}) is 
qualitatively the same in both cases. 
Hence it looks likely
that the grain boundary model is in the same universality class as the XY
model. Also, a detailed comparison indicates a faster
convergence  to the
KT behaviour for the lattice grain boundary model
than for the XY model in Villain's formulation.
The same analysis for the XY model in the cosine form
yields non-unique results, since the data of 
\cite{GUPBAI,WOLF} have larger statistical errors.

\subsection{Numerical results at the transition point}

We now come to the simulations near the transition point.
In the following we will concentrate on the fourth-order cumulant
\be
U=1-\frac{\langle \vec{m}^4 \rangle}{3 \langle \vec{m}^2 \rangle ^2}
\en
and the susceptibility. At the transition point 
finite-size scaling (FSS) implies scale invariance
of $U$ and $\chi \sim L^{2-\eta}$ for large enough lattices. 
We use this FSS behaviour to locate $\beta_{\rm c}$. 
For these simulations we used additional lattices of size 
$L^2=36^2$.

\begin{figure}[t]
\vspace{-0mm}
\begin{center}
\centerline{\epsfysize=10.0cm
\epsfxsize=11.0cm
\epsfbox{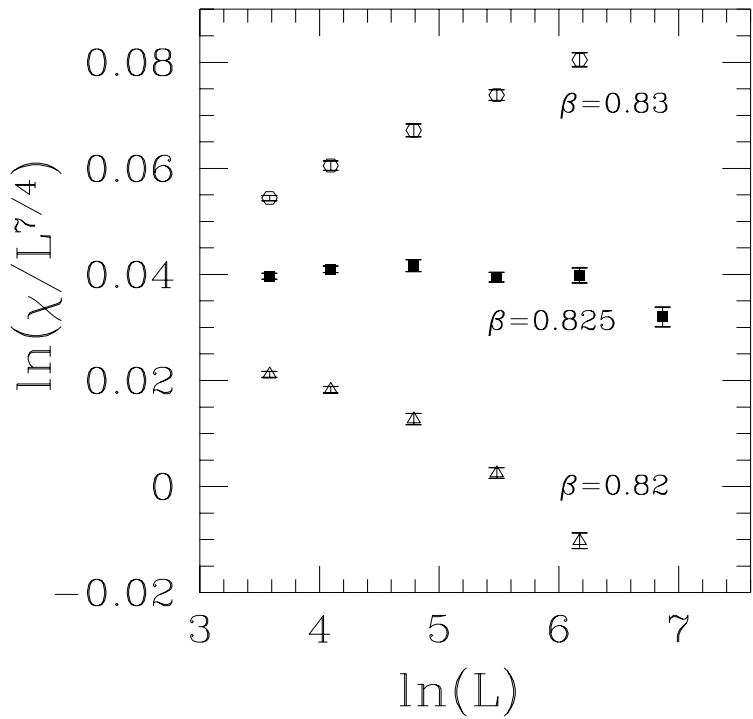}}
\parbox{13.5cm}{\caption{\label{fig_FSSSU}
Finite-size scaling of the susceptibility 
for various inverse temperatures. The slope 
gives the deviation from $\eta=1/4$.
}}
\end{center}
\end{figure}

At the  transition point 
the susceptibility should diverge with the size of the system as
\be
\chi \sim L^{2-\eta} \ .
\en
For $\beta>\beta_{\rm c}$ $\eta$ is a function of the temperature. 
For $\beta$'s below the transition point, one has to take  
corrections for finite correlation lengths of the order ${\cal O}(L/\xi)$
into account. 
In fig.~\ref{fig_FSSSU} we plot $ln(\chi/L^{7/4})$
versus $ln(L)$ for three different $\beta$'s near the transition point.
The slope gives the deviation from $\eta=1/4$. We extracted the values of
$\eta$ from linear fits of the asymptotic behaviour. Our results
are collected in table \ref{tab_eta}.
\begin{table}[b]                                                                
\begin{center}                                                                  
\parbox{13.5cm}{\caption{ \label{tab_eta}                                
The exponent $\eta$ for different $\beta$'s as obtained from FSS.
}}                                                                              
\end{center}  
                
\begin{center}                     
\begin{tabular}{|r@{.}l||*{2}{r@{.}l|}c|}                      
\hline                                                                          
\multicolumn{2}{|c||}{$\beta$}   & \multicolumn{2}{c|}{$\eta$} 
&   \multicolumn{2}{c|}{$\chi^2$/dof} &  
used lattices \\                               
\hline\hline                                                                    
 0&{82}    &  0&{$2663(13)$}     &   0&{89}   & $L = 120-480$    \\
\hline                                                                          
 0&{825}   &  0&{$2535(10)$}        &  2&{91}   &   $L= 120-960$  \\
\hline                                                       
 0&{83}    &  0&{$2404(7)$}     &   $ 0$&{01}   &  $L= 60 - 480$  \\
 \hline                                  
 0&{84}    &  0&{$2242(13)$}      &  1&{49}   & $L = 60-240$    \\
\hline                                                                          
\end{tabular}  
\end{center}                                                                    
\end{table}                                                                     
Obviously, a requirement of  
$\eta=0.25$ yields $\beta_{\rm c} \approx 0.826$. Since there
is a tendency  of the slope to decrease with larger lattices, the
real scale invariance may take place at larger inverse temperatures.
On the other hand, with $\beta_{\rm c,KT3} \approx 0.836$ we got an
$\eta$ which is about  \mbox{$8\,\%$} below the theoretical value.
Again this situation is similar as in the XY model \cite{GUPBAI,JANNAT}.
\begin{figure}[t]
\vspace{-0mm}
\begin{center}
\centerline{\epsfysize=10.0cm
\epsfxsize=11.0cm
\epsfbox{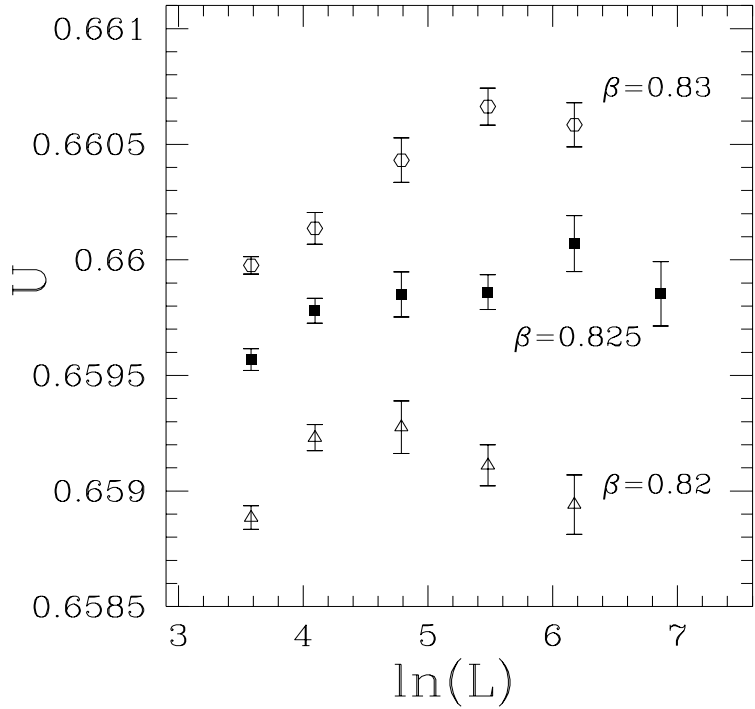}}
\parbox{13.5cm}{\caption{\label{fig_FSSCU}
Finite-size scaling of the cumulant in the
vicinity of the transition point.
}}
\end{center}
\end{figure}

Our results for the dependence of $U$ on the lattice size
are shown in fig.~\ref{fig_FSSCU}.
Apart from the fact that the statistical errors are larger,
we get  a similar estimate as before.   
Also the slopes of $ln(\chi/L^{7/4})$ and of $U$ both
decrease with increasing $L$. Therefore, the value of $\beta_{\rm c}$
can be taken as a lower bound of the large $L$ limit. If we
compare this value with those of the KT and PL fits 
(eqn.~(\ref{betacKT}) and (\ref{betacPL}))  we find that $\beta_{\rm c,KT}$
is in better agreement than $\beta_{\rm c,PL}$. This is an
additional hint favouring a KT transition.

\subsection{Logarithmic corrections to KT scaling laws}

Assuming an exponential divergence as predicted by  Kosterlitz
and Thouless, one can ask for corrections of the scaling behaviour of
$\xi$ and $\chi$. The renormalization group analysis of KT
yields the following multiplicative corrections to eqn.~(\ref{chixi}):
\be
\chi  &\sim& t^r \, \xi^{2-\eta} \ , \ph r=-\frac{1}{16} \ .
\en
In the following we will analyze the data under
the assumption that \mbox{$\nu=1/2$}, \mbox{$\eta=1/4$} and 
determine if logarithmical corrections can explain 
the appearing discrepancies from the leading scaling behaviour.
\begin{figure}[t]
\vspace{-0mm}
\begin{center}
\centerline{\epsfysize=9.5cm
\epsfxsize=10.0cm
\epsfbox{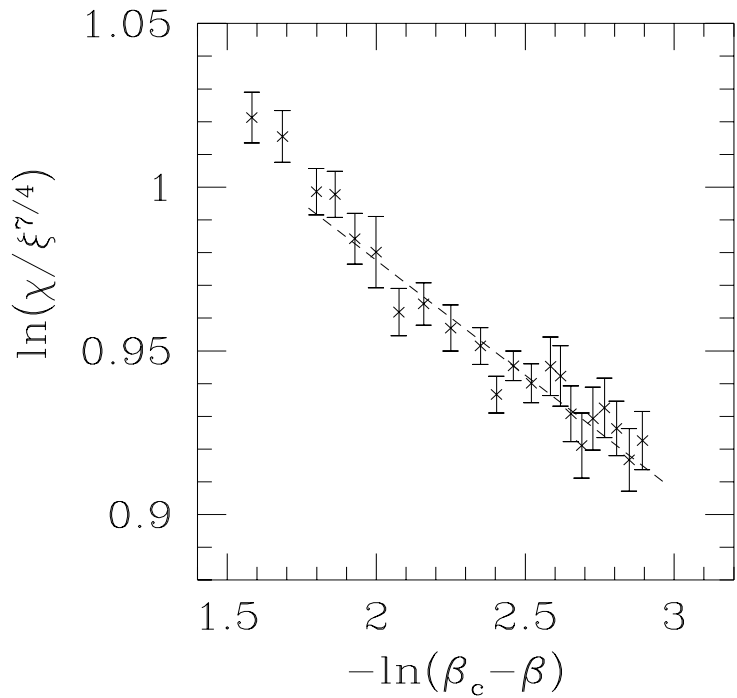}}

\parbox{13.5cm}{\caption{\label{fig_r1}
Logarithmic corrections to  the scaling
of $\chi \sim \xi^{7/4}$. The slope yields $-r$.
The dashed line results from a fit of all 
points with $\xi>6$, i.e.\ 
$-ln(\beta_{\rm c}-\beta)>1.77$.
}}
\end{center}
\end{figure}
Therefore in fig.~\ref{fig_r1}  we plotted 
$ln(\chi/\xi^{7/4})$ as a function of
\mbox{$-ln(t)=-ln(\beta_{\rm c}-\beta)$}, 
where we have taken $\beta_{\rm c}=0.8354$.
The data are consistent with a straight line 
assumption, but  from the slope (for the points $\xi>6$)
we obtained 
\be
r=0.070(5) \ ,
\en
which is completely different from the theoretical value. 
\begin{table} 
\begin{center}                      
\parbox{13.5cm}{\caption{ \label{tab_r}                                
Values of the logarithmic correction exponent $r$
obtained from different methods.
}}                                                                              
\end{center}                  
\begin{center}  

\begin{tabular}{|l|c||*{2}{r@{.}l|}c|}
\hline
\multicolumn{2}{|c||}{method}   & \multicolumn{2}{c|}{$r$} &
\multicolumn{2}{c|}{$\chi^2$/dof} &  points \\
\hline\hline  
$\ln(\chi/\xi^{7/4})$ vs.\ $\ln(\beta_{\rm c}-\beta)$
&  $\xi>6$  & 0&{070(5)} &  0&{86}  & 20 \\
\cline{2-7} 
  &  $\xi>12$  & 0&{056(9)} &  0&{57} & 15 \\
\hline 
$\ln(\chi/\xi^{7/4})$ vs.\ $\ln(\ln(\xi))$
&  $\xi>6$  & 0&{043(3)} &  0&{67} & 20 \\
\cline{2-7} 
  &  $\xi>12$  & 0&{036(6)} &  0&{56} & 15  \\
\hline 
$\ln(\chi/L^{7/4})$ vs.\ $\ln(\ln(L))$ 
&  $\beta=0.83$  & -0&{0233(10)} &  0&{26} & 5 \\
\cline{2-7} 
&  $\beta=0.84$  & -0&{059(4)} &  4&{20} & 3 \\ 
\hline 
\end{tabular}

\end{center}                                                                    
\end{table}                                                                     
For the interval  $\xi>12$ we got \mbox{$r=0.056(9)$}.
This means that the value of $r$ decreases, so that we
probably have not reached the scaling region.
On the other hand a
different $\beta_{\rm c}$ results in only small changes
of $r$.

Alternatively 
one can also try to replace $-ln(t)$ by $2\,ln(ln(\xi))$ 
in fig.~\ref{fig_r1},
as it follows from eqn.~(\ref{KTxi}). This has the advantage, 
that no information about 
$\beta_{\rm c}$ is needed. It is clear that the relation 
is only fulfilled in the limit $t \rightarrow 0$,
where $b_\xi\, t^{-\nu} \gg |ln(a_\xi)|$.
For the actual values of $\beta$ this is not the case, as  can 
be calculated using the parameters of table~\ref{KTvsPL}. 
Nevertheless one gets a linear behaviour which leads to  
$r=0.043(3)$ ($\xi>6$) and $r=0.036(6)$ ($\xi>12$), respectively. 
All data are summarized in table \ref{tab_r}.

A different approach to estimate $r$ is based one FSS.
\begin{figure}[t]
\vspace{-0mm}
\begin{center}
\centerline{\epsfysize=9.5cm
\epsfxsize=10.0cm
\epsfbox{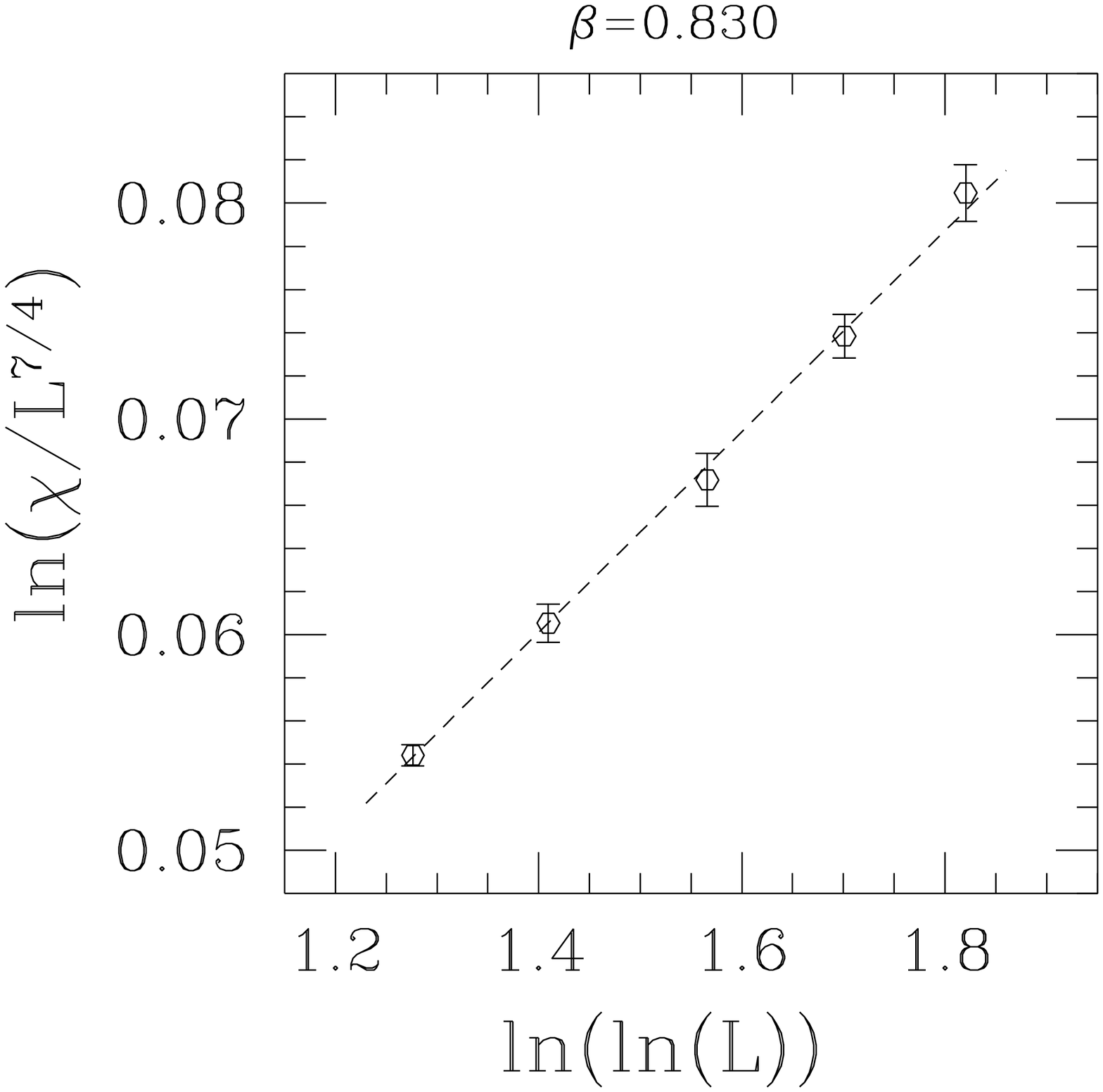}}
\parbox{13.5cm}{\caption{\label{fig_r2}
Logarithmic corrections to the finite-size scaling of the susceptibility
near $\beta_{\rm c}$. We derived $r$ from the slope $-2r=0.0466\,(20)$. 
}}
\end{center}
\end{figure}
At the transition point \mbox{($\xi \gg L$)} one expects
\be
\chi  &\sim& (\ln(L))^{-2r} \, L^{2-\eta}  \ .
\en
In fig.~\ref{fig_r2} we show the data at $\beta=0.83 \approx \beta_{\rm c}$. 
Indeed we can 
observe the predicted linear behaviour. From the slope we obtain
\be
r=-0.0233 \,(10) \ .
\en
Although the value is closer to the theoretical one, it is still in
disagreement with it. A higher (lower) value of $\beta$ would result in a
decrease (increase) of $r$. At $\beta=0.84$ we got $r=-0.059(4)$.

The analysis for multiplicative corrections in the XY model
yields similar results \cite{JANKE,KENIRV}, i.e.\ 
positive values of the order ${\cal O}(0.05)$ from the data
of the high-temperature phase and a negative value of
approximate $-0.03$ from finite-size scaling.


\section{Conclusions}

We presented a detailed Monte Carlo study of a two-dimensional grain 
boundary model on the lattice. 
The investigations were performed in the high-temperature phase
and near the phase transition point.

The behaviour of the specific heat  as a function of 
the inverse temperature $\beta$ 
was characterized 
by a smooth peak at an inverse temperature below the critical value. 
Moreover for 
lattices of $L=120$ or larger no finite-size effects were seen. 
This is consistent 
with the KT scenario.

In the high-temperature phase we examined the dependency of the 
correlation length and susceptibility on $\beta$ and $T$. We showed
that the data are always in agreement  with a KT-like divergence.
The critical exponent $\nu$ was estimated as $\nu=0.46(7)$. In all cases 
fits with a power-law divergence are characterized by a larger $\chi^2$/dof\
than the corresponding KT fit. The exponent $\eta$ was derived 
from the relation 
of $\xi$ and $\chi$ for $\beta \rightarrow \beta_{\rm c}$. 
We got $\eta=0.270(3)$.

The simulations in the vicinity of the transition point were 
used to measure the
finite-size scaling of the fourth-order cumulant and susceptibility.
The resulting estimates of $\beta_{\rm c}$ are closer 
to the value from the KT fits than to the estimate from 
the PL fits. Deviations may be explained by the fact that the lattices used
are still too small.

Additionally we discussed an attempt to explain differences of the critical
exponent $\eta$ from the predicted KT values by multiplicative corrections.
We showed that both finite-size scaling at $\beta_{\rm c}$ and simulations
in the high-temperature phase are consistent with such an ansatz. However the 
values of $r$ are inconsistent and far from the 
theoretical value $r=-1/16$.
As before the reason might be too small lattice sizes.

A comparison of our results showed that the data are in 
accordance with the ones of 
the XY model. The universal critical exponents  
agree within statistical errors, while multiplicative corrections
are qualitatively the same. We take this as evidence
 that both 
models lie in the 
same universality class.

%
%

\end{document}